\documentclass[%
reprint, twocolumn,
superscriptaddress,
showpacs,preprintnumbers,
amsmath,amssymb, aps, pra,
]{revtex4}

\usepackage{graphicx}
\usepackage{dcolumn}
\usepackage{bm}
\usepackage{color} 
\usepackage{CJK}

\def\ra{{\rangle}}

\newcommand{\beq}{\begin{equation}}
\newcommand{\eeq}{\end{equation}}
\newcommand{\beqa}{\begin{eqnarray}}
\newcommand{\eeqa}{\end{eqnarray}}

\begin{document}
\title{Fast and robust population transfer in two-level quantum systems with dephasing noise and/or systematic frequency errors}
\author{Xiao-Jing Lu}
\affiliation{Department of Physics, Shanghai University, 200444
Shanghai, People's Republic of China}
\affiliation{Departamento de Qu\'{\i}mica F\'{\i}sica, UPV/EHU, Apdo 644, 48080 Bilbao, Spain}

\author{Xi Chen}
\affiliation{Department of Physics, Shanghai University, 200444
Shanghai, People's Republic of China}
\affiliation{Departamento de Qu\'{\i}mica F\'{\i}sica, UPV/EHU, Apdo 644, 48080 Bilbao, Spain}
\author{A. Ruschhaupt}
\affiliation{Department of Physics, University College Cork, Cork,
Ireland}

\author{D. Alonso}
\affiliation{Departamento de F\'{\i}sica Fundamental y Experimental,
Electronica y Sistemas and IUdEA, Universidad de La Laguna, 38203 La
Laguna, Spain}

\author{S. Gu\'{e}rin}
\affiliation{Laboratoire Interdisciplinaire Carnot de Bourgogne, CNRS UMR 6303,
Universit\'{e} de Bourgogne, BP 47870, 21078 Dijon, France}

\author{J. G. Muga}
\affiliation{Departamento de Qu\'{\i}mica F\'{\i}sica, UPV/EHU, Apdo
644, 48080 Bilbao, Spain}
\affiliation{Department of Physics, Shanghai University, 200444
Shanghai, People's Republic of China}

\date{\today}
\begin{abstract}
We design, by invariant-based inverse engineering,  driving fields
that invert the population of a
two-level atom in a given time, robustly with respect to dephasing
noise and/or systematic frequency shifts. Without imposing constraints,
optimal protocols are  insensitive to the perturbations but need an infinite
energy. For a constrained value of the Rabi frequency,
a flat $\pi$ pulse is the least sensitive protocol to phase noise but not
to systematic frequency shifts, for which we describe and optimize
a family of protocols.
\end{abstract}
%
%
%
\maketitle
%
\section{Introduction}
The coherent manipulation of quantum systems with time-dependent
interacting fields is a major goal in atomic, molecular and optical
physics, as well as in solid-state devices, for fundamental studies,
Nuclear Magnetic Resonance and other spectroscopic techniques,
metrology, interferometry, or quantum information applications
\cite{Allen,Bergmann,Vitanov-Rev1,Kral,Molmer,Guerin}. Two-level systems
are ubiquitous in these areas,
and the driving of a population inversion is an important operation
that should be typically fast, faithful, stable with respect to
different types of noise and perturbations, and of course  ``feasible in
practice''. The later requirement depends on the specific system but
may be sensibly quantified by setting
constraints on  the possible values of  the control parameters.  These
constraints imply quantum speed limits that
could  be satisfied by optimized protocols.

In a recent paper \cite{Andreas}, the stability of fast population
inversion protocols with respect to amplitude noise and to
systematic perturbations of the driving field was studied, and
optimally stable protocols were found by making use of
invariant-based inverse engineering and perturbation theory.
Our aim here is to extend the analysis to dephasing noise, which may be the dominant source of
decoherence due to environmental effects or the randomly fluctuating
frequency of the control field, and
to systematic frequency errors. By ``systematic error'' we mean
here a constant shift
of the frequency with respect to the one in the
ideal protocol, due e.g. to calibration imperfections or inhomogeneous
broadening.

We shall make use, as in \cite{Andreas,ChenPRA}, of invariant-based
inverse-engineering, which is summarized  in Sec. \ref{inva}.
Section \ref{dn} describes the system and the perturbations
by a Lindblad master
equation. Perturbation theory is then used in Sec. \ref{pn} to derive an expression
for the sensitivity of population inversion with respect to
dephasing noise or systematic frequency errors, and optimal protocols are
defined with or without constraints.
Section \ref{sfe} deals with systematic frequency errors and,
finally, both types of
perturbation -- due to the dephasing noise and constant frequency offset -- are
combined in Sec. \ref{cb}. We shall for concreteness use a language appropriate for two-level atoms in
optical fields, but the results are applicable to other two-level quantum systems.
%
%
%
%
%

\section{Shortcuts to adiabaticity \label{inva}}

\subsection{Dynamical invariants}
We consider a two-level quantum system driven by a time-dependent
Hamiltonian of the form
\begin{equation}
\label{H0} H_{0}(t)=\frac{\hbar}{2}
\left(
\begin{array}{cc}
-\Delta(t) & \Omega (t) \\
\Omega (t) & \Delta(t) \\
\end{array}
\right),
\end{equation}
in the basis $|1\rangle={1\choose 0}$, $|2\rangle={0\choose 1}$.
Eq. (\ref{H0}) corresponds to
a laser-adapted interaction picture,  where the rapid oscillations
of the field have been transformed out, and $\Delta(t)$ and
$\Omega (t)$ are the time-dependent detuning and (real) Rabi
frequencies.
Associated with this time-dependent Hamiltonian there are Hermitian
dynamical invariants $I(t)$, fulfilling $\partial I/\partial
t+(1/i\hbar)[I,H_{0}]=0$, so that their expectation values remain
constant. $I(t)$ may be parameterized as \cite{ChenPRA,spinQD}
\begin{equation}
\label{I}
 I(t)=\frac{\hbar}{2}\Omega_{0}
\left(
                \begin{array}{cc}
                  \cos\theta & \sin\theta e^{-i\beta} \\
                  \sin\theta e^{i\beta} & -\cos\theta
                \end{array}
\right),
\end{equation}
where $\Omega_{0}$ is an arbitrary constant (angular) frequency to
keep $I(t)$ with dimensions of energy, and $\theta\equiv\theta(t)$
and $ \beta\equiv\beta(t)$ are time dependent angles. Using the
invariance condition we find the differential equations
\beqa
\label{thetadot}
\dot{\theta} (t)&=& -\Omega (t)\sin\beta (t), \\
%
\label{betadot}
\dot{\beta}(t)&=&-\Omega (t)\cot\theta (t)\cos\beta(t)-\Delta (t).
\eeqa
The eigenstates of the invariant $I(t)$, satisfy
$I(t)|\phi_{n}(t)\rangle=\lambda_{n}|\phi_{n}(t)\rangle$ ($n=\pm; \lambda_{\pm}=\pm\hbar\Omega_{0}/2$).
Consistently with orthogonality and normalization they can be
written as
\begin{eqnarray}
\label{phit} |\phi_{+}(t)\rangle=\left(\begin{array}{c}
                      \cos\frac{\theta}{2}e^{-i\beta} \\
                      \sin\frac{\theta}{2}
                    \end{array}
                    \right),
|\phi_{-}(t)\rangle=\left(\begin{array}{c}
                      \sin\frac{\theta}{2} \\
                      -\cos\frac{\theta}{2}e^{i\beta}
                    \end{array}
                    \right).
\end{eqnarray}
According to Lewis-Riesenfeld theory \cite{LR}, the solution of the
time-dependent Schr\"{o}dinger equation,
up to a (global) phase factor, can be expressed as
\begin{equation}
\label{psit}
|\Psi(t)\rangle=\Sigma_{n}c_{n}e^{i\gamma_{n}(t)}|\phi_{n}(t)\rangle,
\end{equation}
where the $c_{n}$ 
are time-independent amplitudes, and the $\gamma_{n}(t)$ are
Lewis-Riesenfeld phases
\beq
\gamma_{n}(t)
\equiv \frac{1}{\hbar}\int_{0}^{t}\langle\phi_{n}(t')|i\hbar
\frac{\partial}{\partial t'}-H_{0}(t')|\phi_{n}(t')\rangle dt',
\eeq
where the initial time $t_i$ has been chosen as $t_i=0$.
In our two-level system model, the Lewis-Riesenfeld phases take the form
\beq
\gamma_{\pm}(t) = \pm \frac{1}{2} \int_{0}^{t} \left(
\dot{\beta}+\frac{\dot{\theta}\cot\beta}{\sin\theta} \right) dt' .
\eeq

\subsection{Inverse engineering}
We shall now review briefly the inverse engineering of population
inversion based on dynamical invariants. The initial and final states of the
process are set as $|\Psi(0)\rangle=|2\ra\equiv$$0\choose 1$ and
$|\Psi(T)\rangle=|1\ra\equiv$$1\choose 0$ respectively. The state
trajectory between them may be parameterized according to one of the
eigenstates, $|\phi_{n} (t)\rangle$, of the invariant. By using
$|\phi_{+} (t) \rangle$ in Eq. (\ref{phit}), the boundary conditions
\cite{ChenPRA}
\beq
\label{boundary0}
\theta(0)= \pi,~~~
\theta(T) = 0,
\eeq
guarantee the desired  initial and final states. If in addition
\beq
\label{boundarydot}
\dot{\theta}(0)=0,~~~ \dot{\theta}(T)=0,
\eeq
then $\Omega (0)=\Omega (T) =0$, and $H_{0}(t)$ and $I (t)$ commute at
times $t=0$ and $t=T$. Apart from the boundary conditions,
$\theta(t)$ and  $\beta(t)$ are in principle quite arbitrary, and  the
possible divergences at multiples of $\pi$ of $\beta$ may be
canceled with a vanishing $\dot{\theta}$. The commutativity
at the time boundaries implies that the operators share the eigenstates so, if $H_0 (t)$ remains
constant before and after the process time interval $[0, T]$, then
the initial eigenstates of $H_0(t<0)$ will be smoothly inverted into
final eigenstates of $H_0(t>T)$ following the invariant
eigenvectors. If the condition (\ref{boundarydot}) is not imposed,
the states at $t=0$ and $t=T$ will not be stable (stationary
eigenstates), so a sudden jump is required in the Hamiltonian to
make them so. The flat $\pi$ pulse is a clear simple example, where
the Rabi frequency jumps from or drops to zero abruptly.
Once $\theta(t)$ and $\beta(t)$ have been specified (the interpolation
may be based on simplicity or to satisfy further conditions) the
Rabi frequency and detuning are given, from Eqs. (\ref{thetadot}) and (\ref{betadot}),
by
\beqa
\label{omega}
\Omega (t) &=& -\frac{\dot{\theta}(t)}{\sin\beta(t)},\\
\label{delta}
\Delta(t) &=& \dot{\theta}(t)\cot\beta(t)\cot\theta(t)-\dot{\beta}(t).
\eeqa
For $\beta(t)= \pi/2$,  then $\Delta=0$, and
\beq
\label{pipulse}
\int^{T}_0 \Omega (t)\, dt= \pi
\eeq
corresponds to a $\pi$ pulse. In particular, for $|\dot{\theta}|=
\pi/T$, the flat $\pi$ pulse ($\Omega(t) = \pi/T$ and $\Delta=0$)
minimizes, for a given $T$, the maximal value of $\Omega(t)$ along
the protocol, $\Omega^{\rm max} = \mbox{max}_t |\Omega(t)|$.

\section{Model for dephasing noise and systematic frequency shifts}
\label{dn}%
We assume that the dynamics of the two-level quantum system with
dephasing noise and systematic error may be described by a master
equation in Lindblad form \cite{book,Lindblad},
\begin{equation}
\frac{\partial\rho}{\partial t}=-\frac{i}{\hbar}[H_{0}+
H_{1},\rho]-\frac{1}{2}(\Gamma_{d}^{\dag}
\Gamma_{d}\rho+\rho\Gamma_{d}^{\dag}\Gamma_{d}-2\Gamma_{d}\rho\Gamma_{d}^{\dag}),
\end{equation}
where $\rho$ is the density matrix, $H_{0}$ is the unperturbed
Hamiltonian (\ref{H0}), $H_{1} = \hbar \delta_0 \sigma_z/2$
describes the systematic frequency error ($\delta_0$ is a
constant frequency shift), $\Gamma_{d}=\gamma_{d}\sigma_{z}$ is the
Lindblad operator corresponding to a dephasing rate $2\gamma_d^2$
\cite{Sarandy07}, and $\sigma_z$ is the $z$ Pauli matrix.
This master equation results from averaging over white noise realizations
of the fluctuation of the laser frequency or more generally, of
the detuning, see the appendix
in \cite{Andreas}. The designed detuning thus may generally be
perturbed in our model by a systematic
constant offset and a random contribution with zero mean and
delta-function correlation function.
The dephasing effect corresponds to the randomization of the relative phases of coherent superpositions of states.
It is detrimental for a process of complete population transfer, since the dynamics goes necessarily through a transient superposition
of states. Very few analytic solutions are known for such systems (see for instance \cite{Kyoseva} and the approximative results beyond the exact resonance in Ref. \cite{Lacour}). In the adiabatic context, the effects of dephasing can be reduced by a fast sweeping through the resonance, which however induces nonadiabatic effects. Adiabatic solutions reaching a compromise have been proposed in \cite{Lacour08}. Ideal sudden-switch transitions have been suggested in \cite{Poggi}.
We show below that, for a given peak Rabi frequency, the flat $\pi$-pulse is optimally robust with respect to the dephasing effect.
We next analyze a family of (continuous) pulsed Rabi frequencies which are very close to the optimality of the flat $\pi$-pulse.
It is next considered for a robust process with respect to systematic frequency errors and also combined with the dephasing error.

It is useful to represent the density matrix by the Bloch vector
$\vec{r}(t)=(r_{x},r_{y},r_{z})$,
\begin{equation}
\label{r} \vec{r}(t)=\left(
       \begin{array}{c}
         \rho_{12}+\rho_{21} \\
         i(\rho_{12}-\rho_{21}) \\
         \rho_{11}-\rho_{22} \\
       \end{array}
     \right),
\end{equation}
as $\rho=\frac{1}{2}(1+\vec{r}\cdot\vec{\sigma})$, where
$\vec{\sigma}=(\sigma_{x},\sigma_{y},\sigma_{z})$ is the Pauli
vector. The Bloch equation corresponding to the master equation can
be written as
\begin{equation}
\frac{d}{dt}\vec{r}=(\hat{L}_{0}+\hat{L}_{1}+\hat{L}_{d})\vec{r},
\end{equation}
where
\beqa
\hat{L}_{0}=\left(
\begin{array}{ccc}
0 & \Delta & 0 \\
-\Delta & 0 &- \Omega \\
0 & \Omega & 0 \\
\end{array}
\right), \eeqa
\begin{equation}
 \hat{L}_{1}=\left(
  \begin{array}{ccc}
  0 & -\delta_0 & 0 \\
  \delta_0 & 0 & 0 \\
  0 & 0 & 0 \\
\end{array}
\right),
\end{equation}
and
\beqa \hat{L}_{d}=\left(
\begin{array}{ccc}
-2\gamma_{d}^{2} & 0 & 0 \\
0 & -2\gamma_{d}^{2} & 0 \\
0 & 0 & 0 \\
\end{array}
\right). \eeqa
The probability to find the system in $|1\ra$ at time $t$ is
$P_1(t)=\frac{1}{2}[1+r_{z}(t)]$. In the following, we shall
consider the dephasing term $\hat{L}_{d}$ and the systematic
frequency error $\hat{L}_{1}$ as a perturbation, respectively, and
then study both together.

\section{Phase noise\label{pn}}
In this section we set $\delta_0=0$ and consider only phase
noise as the perturbation.
The unperturbed Bloch vector is written as
\begin{equation}
\vec{r}_{0}(t)=\left(
\begin{array}{c}
\sin\theta \cos\beta \\
\sin\theta \sin\beta \\
\cos\theta \\
\end{array}
\right). \label{rsingle}
\end{equation}
Applying time-dependent perturbation theory,
\beq
r_z(T)\simeq 1+\int_{0}^{T}dt\langle\vec{r}_{0}(t)|
\hat{L}_{d}|\vec{r}_{0}(t)\rangle,
\eeq
which results in
\beqa
\label{fe} P_1(T) \simeq 1- \gamma_{d}^{2}\int_{0}^{T}\sin^{2}\theta dt .
\eeqa
By defining the noise sensitivity as \cite{Andreas}
\beqa \label{defineqn}
q_{N}=-\frac{1}{2}\frac{\partial^{2}P_1(T)}{\partial\gamma_{d}^{2}}\bigg|_{\gamma_{d}=0},
\eeqa
and using Eqs. (\ref{fe}) and (\ref{defineqn}), we have
\begin{equation}
\label{qn0} q_{N}=\int_{0}^{T}\sin^{2}\theta\, dt.
\end{equation}
The smaller the noise sensitivity the more stable the fidelity is
with respect to dephasing noise. According to Eq. (\ref{qn0})
$q_{N}$ is zero when $\theta$ is equal to $0$ or $\pi$. Thus a
sudden jump of $\theta$ from $\pi$ to $0$ will cancel the effect of
dephasing noise. (This is consistent with the
sudden-switch transitions in \cite{Poggi}.)
However, a step function for $\theta$ implies an infinite
Rabi frequency according to Eq. (\ref{omega}), and an infinite
energy. Let us consider a time $t^*$
for which $|\dot{\theta}|$ is maximal. Then we can use Eqs.
(\ref{omega}) and (\ref{qn0}) to establish the following
inequalities:
\beqa
\label{bound} \Omega^{\max} q_{N} &=&
\frac{1}{|\sin\beta(t^*)|}\int_{0}^{T}|\dot{\theta}(t^*)| \sin^{2}
\theta dt,
\nonumber \\
&\geq & \frac{1}{|\sin\beta(t^*)|}
\left|\int_{0}^{T}\dot{\theta}\sin^{2}\theta dt\right|,
\nonumber \\
&\geq& \frac{\pi}{2|\sin\beta(t^*)|}\geq\pi/2.
\eeqa
This is a significant relation that sets in particular a lower bound
for the sensitivity when $\Omega^{\max}$ cannot
exceed some predetermined fixed value,
$\Omega^{\max}\leq\Omega^{M}$, due to a finite laser power, or
to avoid multiphoton excitation of other transitions that remain
negligible for weak fields \cite{Zhang}.

A flat $\pi$ pulse with $\beta=\pi/2$ and $\theta = \pi (T-t) / T$,
saturates the bound since
\beq
\Omega = \pi/T, ~~ q_N=T/2.
\eeq

Let us now consider a continuous $\Omega(t)$ based on a $\theta(t)$ function
that satisfies the boundary conditions (\ref{boundary0}) and
(\ref{boundarydot}).
A simple example is
\begin{equation}
\label{theta2} \theta (t)=\left\{
  \begin{array}{ll}
    \pi,  &~~   0 \leq t \leq t_1 \\
    \frac{\pi}{2}\left\{1-\sin\left[\frac{\pi(2 t-T)}{2 WT}\right]\right\}, &~~  t_1 \leq t \leq t_2 \\
    0,  &~~ t_2 \leq t \leq T
  \end{array}
\right.,
\end{equation}
with $t_1= (1-W) T/2$, $t_2= (1+W)T/2$ and $0<W\leq 1$.
From Eq. (\ref{theta2}) and for $t_1 \leq t \leq t_2$,
\begin{equation}
|\dot{\theta} (t)|= \frac{\pi^{2}}{2W
T}\cos\left[\frac{\pi(t-T/2)}{WT}\right] \leq \frac{\pi^{2}}{2WT}.
\end{equation}
We set $\beta=\pi/2$ such that Eq. (\ref{delta}) gives $\Delta=0$.
From  Eq. (\ref{omega}) and for $t_1 \leq t \leq t_2$, we get
\beq
\label{smoothOmega}
\Omega (t) =
\frac{\pi^{2}}{2W T}\cos\left[\frac{\pi(t-T/2)}{WT}\right],
\eeq
with $\Omega (t_1) = \Omega (t_2) = 0$. The maximal value at
$t=T/2$ is
\beq
\label{max0} \Omega^{\max} = \frac{\pi^2}{2WT}.
\eeq
These are (non-flat) $\pi$ pulses satisfying Eq. (\ref{pipulse}).

In the noiseless limit, Eq. (\ref{theta2}) provides complete population inversion
for every $W$ with $0<W\leq 1$. The noise sensitivity, defined by Eq.
(\ref{qn0}), becomes
\beq
q_N=[1+J_0(\pi)]TW/2,
\eeq
where $J_0$ is the Bessel function of the first kind.
This gives $\Omega^{\max}q_N=\pi^2[1+J_0(\pi)]/4\approx 1.7167>\pi/2
\approx1.5708$, for all $T$ and allowed $W$,
only slightly above the bound satisfied by the flat $\pi$ pulse.

\section{Systematic frequency errors
\label{sfe}}
In this section, we shall discuss solely systematic frequency errors described by
$H_1= \hbar \delta_0 \sigma_z/2$ assuming  $\gamma_d=0$.
By using perturbation theory, we obtain
\beqa
|\psi(T) \rangle
&\simeq& |\psi_0(T)\rangle-\frac{i\delta_0}{2}\int_{0}^{T}dt U_0(T,t)\sigma_z|\psi_0(t)\rangle - \left(\frac{\delta_0}{2}\right)^{\!2}
\nonumber \\
&\times& \int_0^T\!\!\!dt\!\! \int_0^t\! dt' U_0(T,t)\sigma_zU_0(t,t')\sigma_z|\psi_0(t')\rangle+...,
\nonumber \\
\eeqa
where $ U_0(T,t) =
|\psi_0(T)\rangle
\langle\psi_0(t)|+|\psi_ \perp(T)\rangle\langle\psi_\perp(t)| $,
$|\psi_{0}(t)\rangle = e^{i\gamma_+} |\phi_{+}(t) \ra $, and
$|\psi_{\perp}(t)\rangle =  e^{i\gamma_-} |\phi_-(t) \ra $. The
probability to find the ground state at $t=T$ is
\beqa
P_1(T) \simeq
1-\left(\frac{\delta_0}{2}\right)^2\left|\int_{0}^{T}dt\langle\psi_\perp(t)|\sigma_z|\psi_0(t)\rangle\right|^2.~~
\eeqa
By defining the systematic error sensitivity as
\begin{equation}
q_S=-\frac{1}{2}\frac{\partial^{2}P_1(T)}{\partial\delta_0^{2}}\bigg|_{\delta_0=0},
\end{equation}
we have
\beq \label{qndelta}
q_S =
\frac{1}{4}\left|\int_0^Tdt\sin\theta e^{im(t)}\right|^2,
\eeq
with
\beq \label{m}
m(t)=2\gamma_+(t)-\beta(t).
\eeq
For example, a flat $\pi$ pulse ($\Omega=\pi/T$, $\theta = \pi(T - t)/T$, and $\beta = \pi/2$) gives
\beq
q_S=(T/\pi)^2. 
\eeq
Note that $q_S$ and $q_N$ have different dimensions.
A major difference between the two types of perturbation
is that there are protocols that nullify $q_S$ without requiring an infinite
$\Omega^{\max}$.
According to Eq. (\ref{qndelta}) a sudden jump from $\pi$ to $0$
leads to a systematic error sensitivity $q_S=0$.
However, as
mentioned before, the sudden transition requires an infinite laser
intensity. To keep $\theta$ continuous and nullify
$q_S$ we may assume -motivated by \cite{Andreas2013}-
\beq\label{mt}
m(t)=2\theta+2\alpha \sin(2\theta),
\eeq
where $\alpha$ is a free parameter, which will be
varied to achieve $q_S=0$.
Setting Eqs. (\ref{mt}) and (\ref{m}) to be equal and doing the time
derivative, we obtain
\begin{equation}
\label{sbeta}
\beta(t)=\cos^{-1}\left(\frac{2M \sin\theta}{\sqrt{1+4M^2\sin^2\theta}}\right).
\end{equation}
with $M= 1+2\alpha\cos(2\theta)$.
Let us calculate the corresponding physical quantities.
Substituting
Eq. (\ref{sbeta}) into Eqs. (\ref{omega}) and
(\ref{delta}), we get for $t_1 \le t \le t_2$
\beqa
\label{somega}
\Omega (t) &=&
-\dot{\theta}\sqrt{1+4M^2\sin^2\theta},\\
\label{sdelta}
\Delta (t) &=& 2 \dot{\theta} \cos\theta \left[ M + \frac{1-4\alpha + 6 \alpha
\cos(2\theta)}{1+4M^2
  \sin^2\theta}
\right].
\eeqa

\begin{figure}[]
\scalebox{0.5}[0.5]{\includegraphics{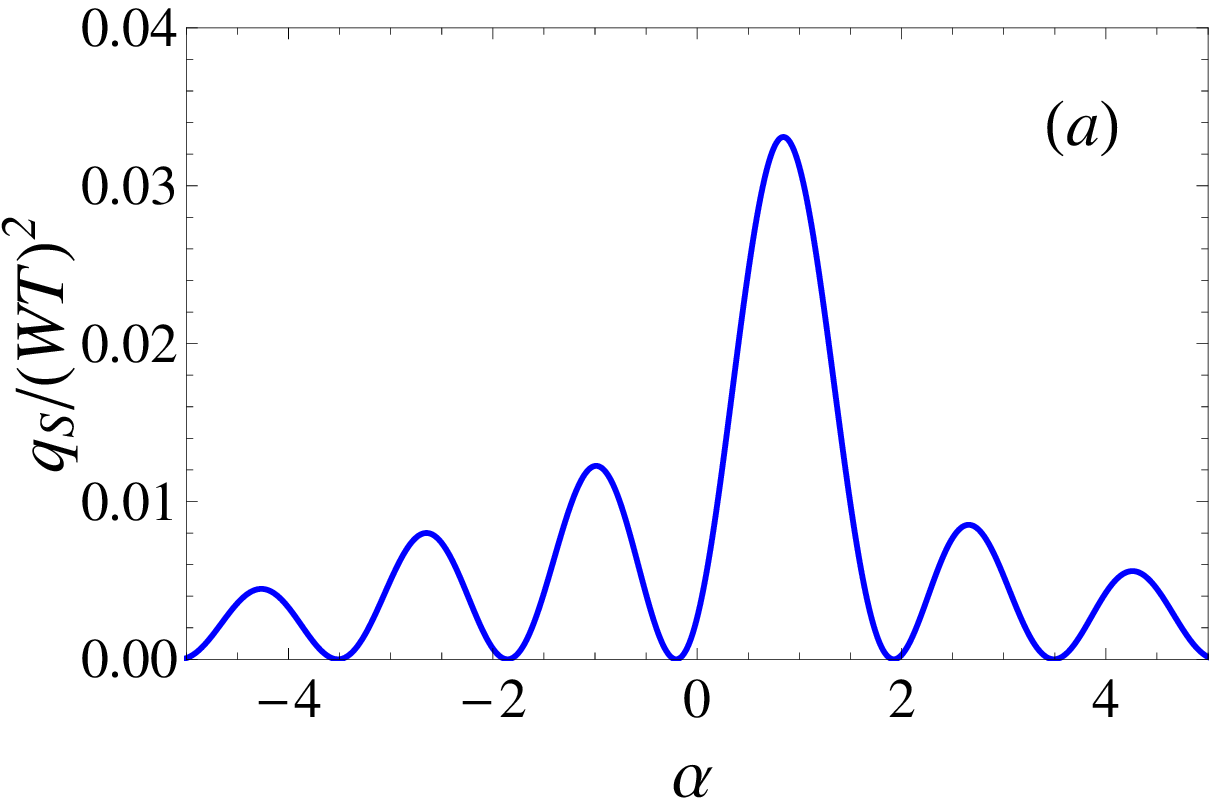}}
\scalebox{0.5}[0.5]{\includegraphics{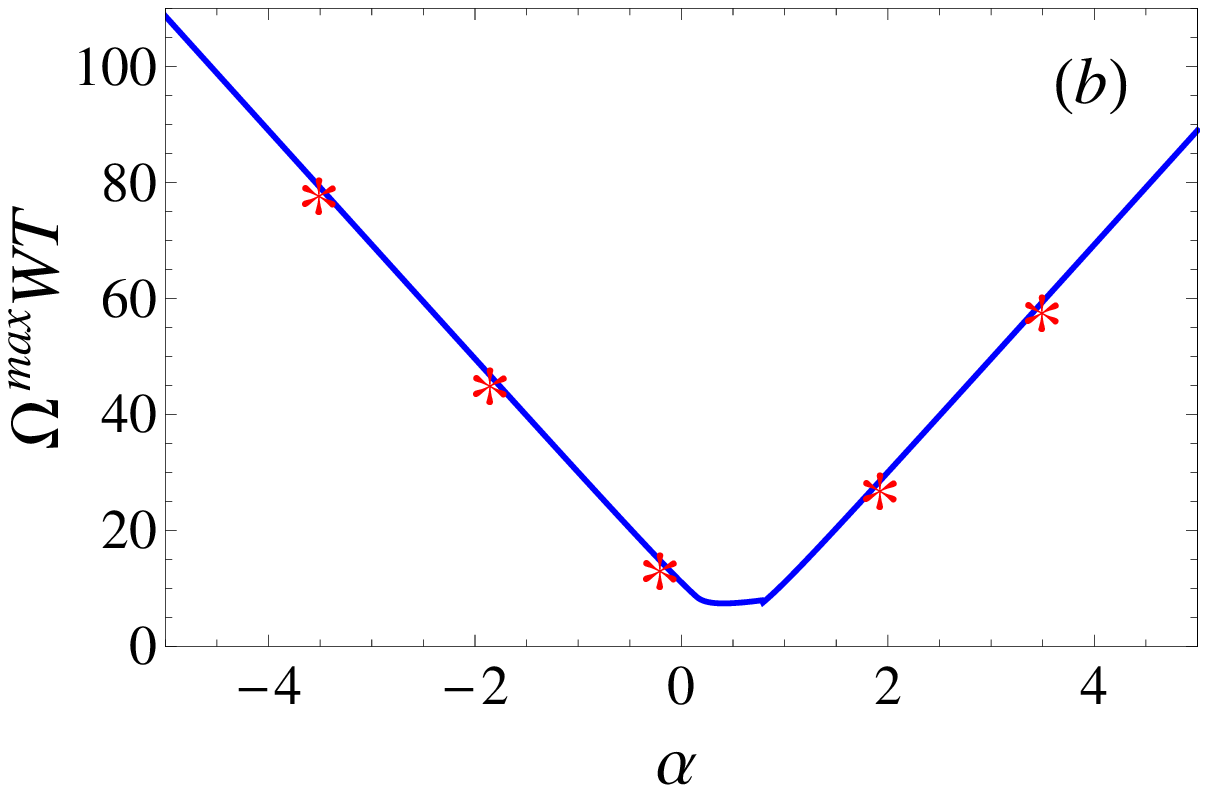}}
\scalebox{0.5}[0.5]{\includegraphics{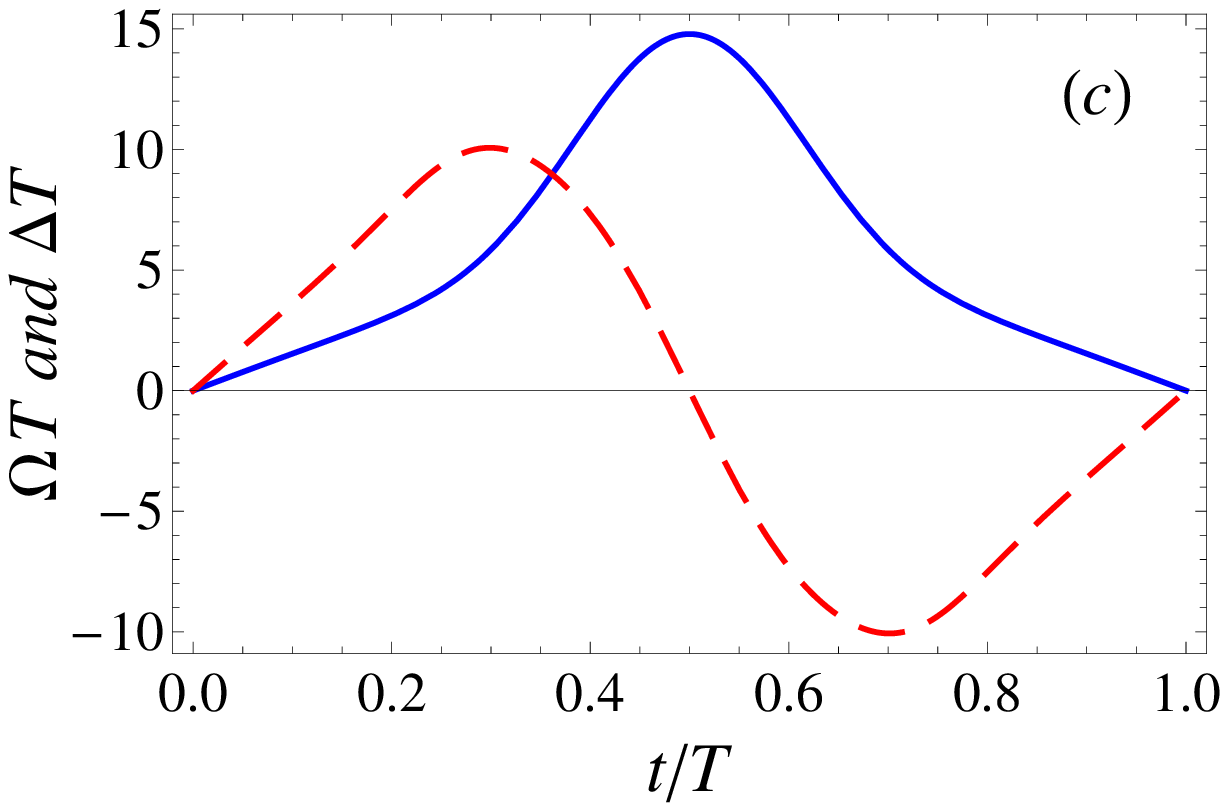}} \caption{(Color
online) (a) Systematic error sensitivity $q_S/(WT)^2$, Eq. (\ref{qS}), and (b) the
Rabi frequency $\Omega^{\max}WT$, Eq. (\ref{omegamax}), versus $\alpha$, where
the stars correspond to $\alpha$ with $q_S =0$.
The coordinate of the start with the minimal $|\alpha|$ with $q_S = 0$ is
$(-0.206, 14.784)$.
(c) Rabi frequency $\Omega T$ (solid blue)
and detuning $\Delta T$ (dotted red) versus time $t/T$,
from Eqs. (\ref{theta2}), (\ref{sbeta}), (\ref{somega}) and (\ref{sdelta}) with
$W=1$, $\alpha=-0.206$. \label{fig1}}
\end{figure}

Now we choose the
$\theta$ as in Eq. (\ref{theta2}).
The systematic error sensitivity is then given by
\beq q_S =
\frac{1}{4}\left|\int_{t_1}^{t_2}dt\sin\theta \exp\left(2i\theta +
2i \alpha \sin(2\theta)\right)\right|^2. \eeq
 Let $t \equiv
\frac{T}{2}(1+\lambda W)$, i.e. $\lambda = (2t-T)/(WT)$, then we get
\begin{eqnarray*}
\lefteqn{q_S = (WT)^2 \Big| \frac{1}{4} \int_{-1}^1 d\lambda \,
\cos\left[\frac{\pi}{2}\sin\left(\frac{\pi}{2}\lambda\right)\right] }
&& \\ &&\times \exp\left\{-i\pi \sin\left(\frac{\pi}{2}\lambda\right) + 2 i
\alpha
\sin\left[\pi\sin\left(\frac{\pi}{2}\lambda\right)\right]\right\}
\Big|^2.
\end{eqnarray*}
This can be simplified further by doing the additional variable
transformation $z=\sin\left(\pi\lambda/2\right)$,
\beq
\label{qS}
q_S = \frac{(WT)^2}{2\pi}\left|\int_{-1}^1 dz\,
\frac{\cos\left(\frac{\pi z}{2}\right)}{\sqrt{1-z^2}} e^{-i\pi z + 2
i \alpha \sin\left(\pi z\right)} \right|^2.
\eeq
The important point is that $q_S/(WT)^2$ is independent of $T$ and
of $W$ and only depends on $\alpha$. This function is shown in Fig.
\ref{fig1}(a). The goal is to choose a value of $\alpha$ such that
$q_S/(WT)^2=0$. The corresponding Rabi frequency is for $t_1 \le t
\le t_2$ and for all $\alpha$
\beq
\label{omx}
\Omega (t) = \frac{\pi^2 \sqrt{1-z^2}}{2 WT}
\sqrt{1+4[1-2\alpha\cos(\pi z)]^2\cos^2\left(\frac{\pi z}{2}\right)
},
\eeq
where
$z=\sin\left(\pi\lambda/2\right)=\sin\left[\pi(2t-T)/(2WT)\right]$
($-1 \le z \le 1$) as defined above and $\Omega(t) = 0$ otherwise.

We are interested in a protocol with $\Omega^{\rm max}$ as small
as possible and therefore an $|\alpha|$ as small as possible. The
value $\Omega^{\rm max} WT$ versus $\alpha$ is also shown in Fig.
1(b). Note that $\Omega^{\rm max} WT$ is independent of $T$ and
$W$ as it can be seen from Eq. (\ref{omx}). The $\alpha$ with the
smallest magnitude fulfilling $q_S=0$ is $\alpha=-0.206$. This value
of $\alpha$ makes the systematic error sensitivity zero for all $W$
and all $T$. For $\alpha <0$, the maximal value of the Rabi frequency at $t=T/2$ is
given by
\beq
\label{omegamax}
\Omega^{\max} = \frac{\pi^2}{2 WT}
\sqrt{1+4(1+2|\alpha|)^2},
\eeq
which increases monotonously with $|\alpha|$. When $\alpha=-0.206$,
$\Omega^{\max}WT= 14.784$ and $q_S =0$, see Fig. \ref{fig1} (a) and (b).

Figure \ref{fig1} (c) represents the Rabi frequency $\Omega(t) T$
and detuning $\Delta(t) T$ versus $t/T$ for $\alpha=-0.206$ and
$W=1$. Both functions are continuous and easy to implement.

%
%

\begin{figure}[]
\scalebox{0.5}[0.5]{\includegraphics{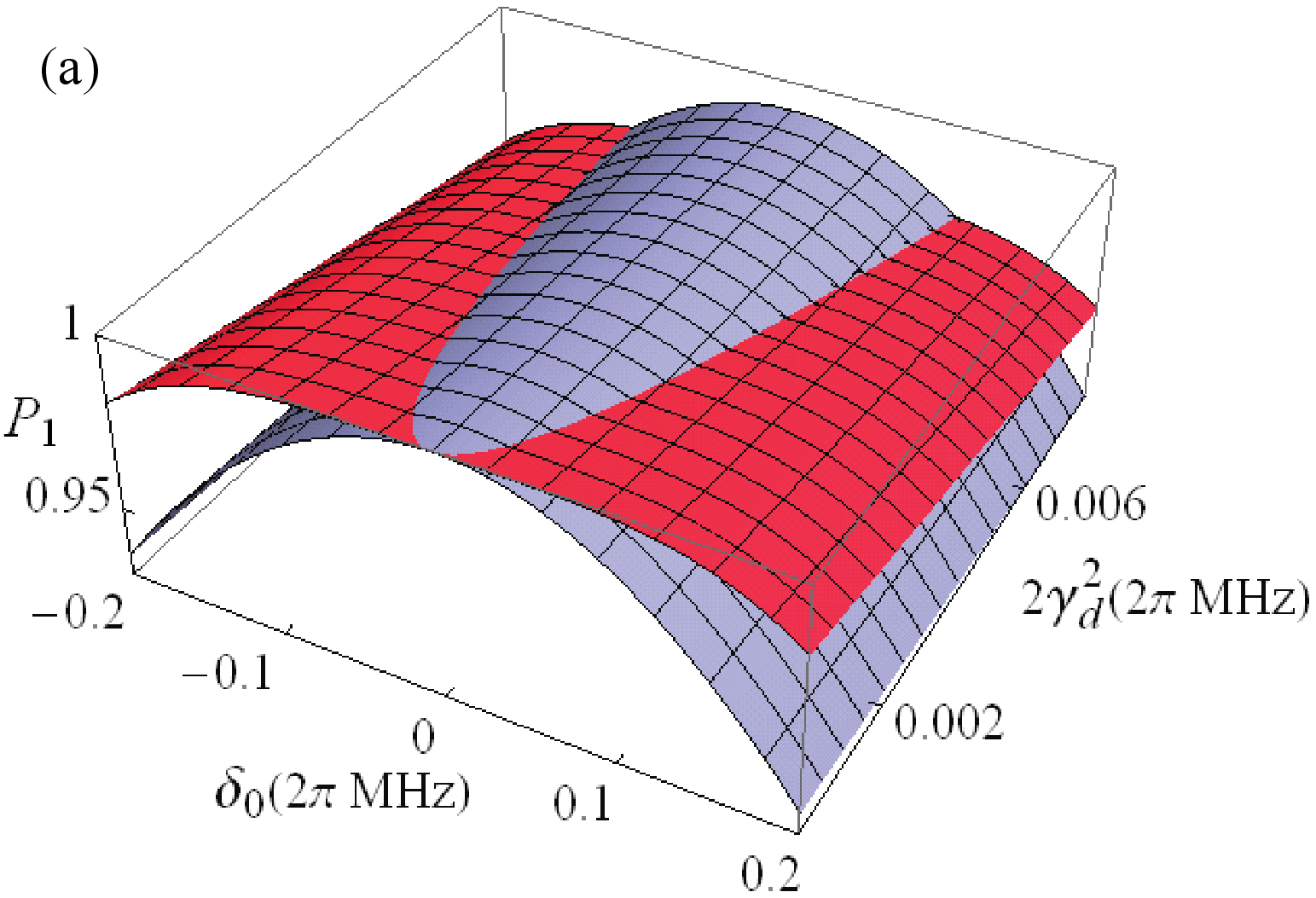}}
\scalebox{0.48}[0.48]{\includegraphics{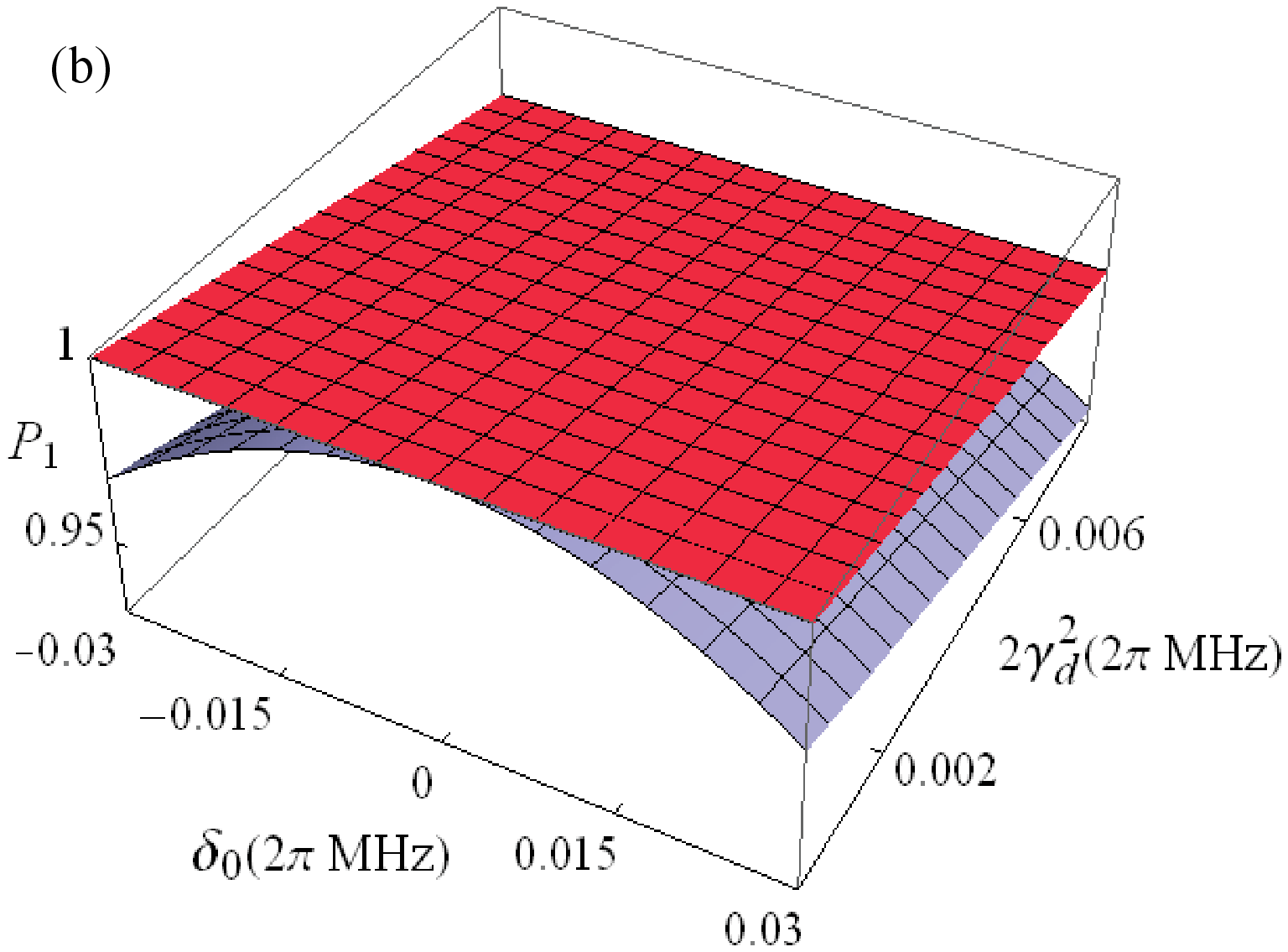}} \caption{(Color
online) Comparison of inversion probability $P_{1}(T)$ for two
protocols. In (a) they share the same maximal Rabi frequency
$\Omega^{\max}=0.784$ $\times$ 2$\pi$ MHz: one is a flat
$\pi$-pulse, optimal for dephasing noise (light blue surface,
prominent around $\delta_0=0$), with
$T_{\pi}=\pi/\Omega^{\max}=0.638$ $\mu$s; the other one (dark red
surface, more prominent around $\gamma_d=0$) has been optimized with
respect to systematic frequency errors within the family described
by Eqs. (\ref{theta2}) and (\ref{sbeta}) with $T =3$ $\mu$s,
$\alpha=-0.206$, and $W=1$. In (b) this later pulse remains the same
but the $\pi$ pulse spans also $3$ $\mu$s, so that its
$\Omega^{\max}=0.167 \times 2 \pi$ MHz.} \label{figcompare}
\end{figure}

\section{Combined perturbations\label{cb}}
Finally, we will consider both types of perturbations (noise and
systematic error) together, so that $P_1(T)\approx 1-\gamma_d^2
q_N-\delta_0^2 q_S$. The best protocol in this case depends on the
relative importance between  dephasing noise and systematic error.
Fig. \ref{figcompare} (a) depicts the final population $P_1(T)$
versus dephasing noise and systematic error perturbative parameters
for two protocols that share the same $\Omega^{\max}$. The first
one is  a flat $\pi$ pulse (light blue surface), which is optimal
with respect to dephasing noise ($q_N=T_\pi/2$, $q_S=(T_\pi/\pi)^2$,
$\Omega=\Omega^{\max}=\pi/T_\pi$), and the second one (dark red
surface) is described in the previous section ($\alpha=-0.206$,
$q_N=[1+J_0(\pi)]TW/2$, $q_S=0$, $\Omega^{\max}= 14.784/WT$. We
choose $W=1$, $T=3$ $\mu$s, and $T_\pi= 0.638$ $\mu$s so that
$\Omega^{\max}$ takes the same value for both protocols. The $\pi$
pulse is the most stable when dephasing noise is dominant whereas
the protocol that nullifies $q_S$ outperforms the $\pi$ pulse
otherwise. In Fig. \ref{figcompare} (b) the $\pi$ pulse is modified
to span also $T =3$ $\mu$s. This lowers its $\Omega^{\max}$ as
well as its robustness.

\section{Discussion}
The design of fast and robust protocols  for coherent population or
state control of a quantum system depends strongly on the type of
noise and/or perturbation. In a previous publication we designed, for the population inversion of a
two-level atom in an electric field, driving fields  which are robust
with respect to amplitude noise or/and systematic
perturbations of the Rabi frequency \cite{Andreas}. Here we have
considered instead excitation frequency shifts with constant offset and/or a white
noise component that generates dephasing.
When the Rabi
frequency is not allowed to increase beyond a certain value,
a flat $\pi$-pulse is the most robust approach
versus phase noise but not with respect to systematic frequency shifts.
The effect of systematic frequency shifts can be minimized
(achieving zero sensitivity) with an alternative family of protocols.
The results obtained here and in \cite{Andreas} indicate that the standard
claim that ``adiabatic methods are robust whereas resonant $\pi$ pulses are not''
does not apply to all possible perturbations. In other words, ``robustness''
is a relative concept. A protocol may be robust with respect to a particular
perturbation but not to others. Depending on the physical conditions, it may be
possible to nullify the sensitivity with respect to different perturbations simultaneously \cite{Andreas2013}.
In the case of phase noise and frequency errors, only the sensitivity with respect to
systematic frequency shifts   can be nullified with finite energy.

The present techniques may as well be applied to find
robust protocols for other
perturbations and decoherence effects including
spontaneous decay and bit-flip \cite{Sarandy07}, with applications in different
quantum systems such as quantum dots \cite{spinQD}, Bose-Einstein
condensates in accelerated optical lattices \cite{Oliver}, or quantum refrigerators
\cite{Tova}.
Combining invariant-based engineering with optimal control techniques \cite{Gorman}
will allow for further stability with different physical constraints.
This work may as well be generalized to consider colored phase noise and
non-Markovian dephasing \cite{Yu,Huelga,Diosi1998,Yu-2,Guerin11,Vega}, as well as alternative phase noise sources and master equations \cite{Poggi}.

\section*{Acknowledgment}
We are grateful to R. Kosloff and Y. Ban for useful discussions.
This work was supported by the National Natural Science Foundation
of China (Grant No. 61176118), Shanghai Rising-Star Program (Grant
No. 12QH1400800), the Basque Country Government (Grant No.
IT472-10), Ministerio de Econom\'{i}a y Competitividad (Grant No.
FIS2012-36673-C03-01), the  UPV/EHU program UFI 11/55,
Spanish MICINN (Grant No. FIS2010-19998), and the European Union (FEDER).
%
%
%

\end{document}